\documentclass[twocolumn,showpacs,preprintnumbers]{revtex4}

\usepackage{graphics}
\begin{document}

\title{Island, pit and groove formation in strained heteroepitaxy}

\author{M.T. Lung$^{1}$, Chi-Hang Lam$^{1}$, and Leonard M. Sander$^{2}$}
\affiliation{
$^1$Department of Applied Physics, Hong Kong Polytechnic University,
Hung Hom, Hong Kong, China \\
$^2$Michigan Center for Theoretical Physics,
Department of Physics, Randall Laboratory, University of Michigan, Ann
Arbor, MI 48109-1120, USA }

\date{\today}

\begin{abstract}
  We study the morphological evolution of strained heteroepitaxial films
  using a kinetic Monte Carlo method in three dimensions. The elastic part of
  the problem uses a Green's function method. Isolated
  islands are observed under deposition conditions for deposition
  rates slow compared with intrinsic surface roughening rates. They are
  semi-spherical and truncated conical for high
  and low temperature cases respectively. Annealing of films at high
  temperature leads to the formation of closely packed islands
  consistent with an instability theory. At low temperature, pits form
  via a layer-by-layer nucleation mechanism and subsequently develop
  into grooves.
\end{abstract}

\pacs{68.65.-k, 68.65.Hb, 81.16.Dn, 81.16.Rf}

\maketitle

Epitaxial growth techniques have been used to deposit strained
coherent films on substrates of a different materials with
a mismatched lattice constant. This is called heteroepitaxy.
Many experiments have shown that
beyond a threshold film thickness, an
array of three dimensional (3D) nanosized islands self-assembles
under favorable growth conditions
\cite{Shchukin,Politi,Freund}.  These results are of considerable 
interest since the islands  behave as quantum dots and are expected
to find applications in future microelectronic devices.  The most
intensively studied examples include Ge/Si(100) and more generally its
alloy variant Si$_{1-x}$Ge$_x$/Si(100)
\cite{Mo,Jesson1996,Vailionis,Floro,Sutter}. The island morphology
depends strongly and often non-trivially on the lattice
misfit dictated by the Ge concentration as well as 
growth conditions including temperature and deposition rate. In addition,
other interesting nanostructures including 3D pits, grooves and
quantum dot molecules composed of coupled islands and pits are also
generated under appropriate conditions \cite{Gray2002,Gray2004}.

In this
letter, we report large scale 3D kinetic Monte Carlo
simulations on the morphological evolution of strained layers.
Our simulations generate  morphologies 
very reminiscent of those observed under various
growth or annealing conditions. We should note that the
simulation of strained layers is computationally challenging due to
the long range nature of elastic interactions.  Previous atomistic
simulations are limited to two dimensions (2D)
\cite{Orr,Khor,Lam,Much} or sub-monolayer coverage \cite{Meixner}.
Continuum computations are less difficult but cannot reliably account
for faceted surfaces and fluctuations which are especially
important at the early stage of roughening \cite{Yang,Zhang,Shenoy}.

We model the film and substrate system by a simple cubic lattice of
balls and springs \cite{Orr,Khor,Lam,Meixner}. The substrate consists
of $64 \times 64 \times 64$ atoms. Periodic boundary conditions in
lateral directions and fixed boundary conditions for the bottom layer
are assumed.  The substrate has a lattice constant $a_s=2.72$\AA ~
which gives an atomic density appropriate for crystalline silicon.  The
lattice constant $a_f$ of the film is related to the lattice misfit
$\epsilon=(a_f-a_s)/a_f$.  Nearest neighboring (NN) and next nearest
neighboring (NNN) atoms are directly connected by elastic springs with
force constants $k_1=2 eV/a_s^2$ and $k_{2}=k_1$ respectively. The
elastic couplings of adatoms with the rest of the system are weak and
are completely neglected.  

Our algorithm imposes solid-on-solid conditions with atomic steps
limited to at most one atom high.  Every topmost atom in
the film can hop to a different random topmost site within a
neighborhood of $l \times l$ columns with equal probability. We put $l=33$.
Decreasing the hopping range does not alter our results significantly.
The hopping rate $\Gamma_m$ of a topmost atom $m$ follows an Arrhenius
form
\begin{equation}
\label{rate}
\Gamma_m = 
{R_0}\exp \left[ -\frac{n_{1m} \gamma_{1} + n_{2m} \gamma_{2}
- \Delta E_m - E_0}{k_{B}T}\right]
\end{equation}
Here, $n_{1m}$ and $n_{2m}$ are the number of NN and NNN of atom $m$
respectively while $\gamma_1=0.085eV$ and $\gamma_2=\gamma_1/2$ are
the corresponding bond strengths. The elastic energy of the hopping
atom is denoted by $\Delta E_m$ and will be explained later. Finally,
we put
$E_0=0.415$eV and $R_0=2D_0/(\sigma a_s)^2$ with $D_0=4.1\times
10^{13}\mbox{\AA}^2 s^{-1}$ and $\sigma^2 =l^2/6$.  This gives the appropriate
adatom diffusion coefficient for silicon (100)
\cite{Savage}. Our choice of the ratios $k_1/k_2$ and
$\gamma_1/\gamma_2$ maximizes the isotropy of the system.

The elastic energy, $\Delta E_m$, has to be repeatedly calculated during
a simulation; this dominates the CPU time. $\Delta E_m$ is
defined as the difference in the strain energy $E_s$ of the whole
lattice at mechanical equilibrium when the site is occupied minus that when it is
unoccupied. Calculating $E_s$ requires solving a long-range elasticity
problem to obtain  the atomic positions of every atom in
the film and the substrate. We have found it possible to significantly speed up the calculation by appling an exact Green's function method. A  method of this type was
 introduced by Tewary \cite{Tewary} in the context of point impurities.  We generalized 
 the technique to free surfaces  in Ref. \cite{Lam}. The result of these developments is that we can solve the elastic problem at a  surface site  using  reduced equations
involving only other surface atoms.  Moreover, we use a surface coarsening scheme
in which morphological details of the surface far away from
atom $m$ are averaged \cite{Lam}. As a result, calculating $\Delta
E_m$ involves only about 160 effective particles and takes less than
one second on a 3GHz pentium computer.  Hopping events are then
sampled using an acceptance-rejection algorithm aided by quick
estimates of $\Delta E_m$ which enables a high acceptance probability.
A simulation reported here typically involves $10^6$ successful
hopping events and takes 10 days to complete. We have considered large
misfit and in some cases also high deposition rate so that the
computations can be manageable.

\begin{figure}
\resizebox{0.4 \columnwidth}{!}{\includegraphics{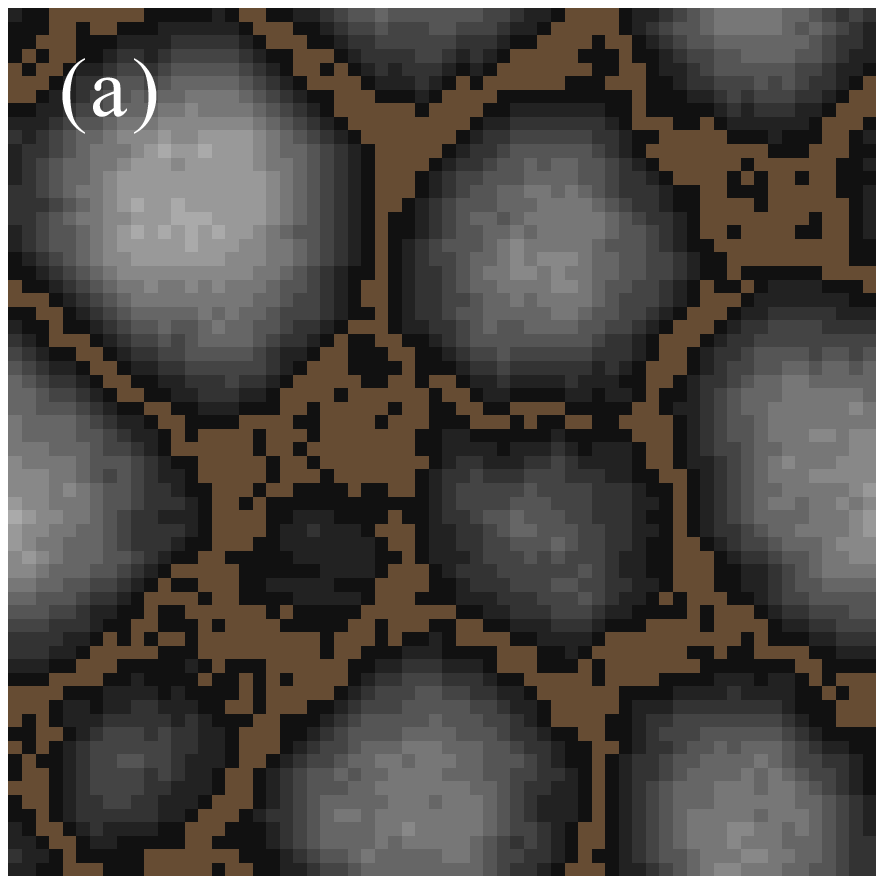}}
~
\resizebox{0.4 \columnwidth}{!}{\includegraphics{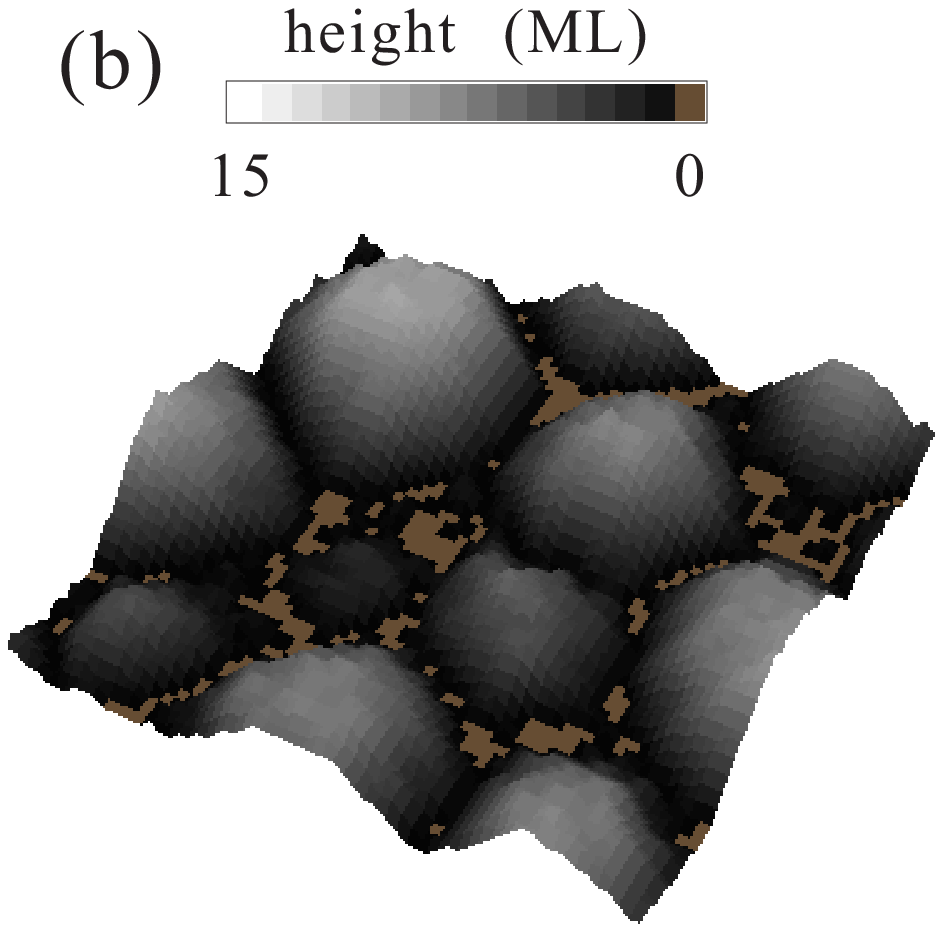}}
\caption{
  \label{F-D1000}
  Surface from simulation of deposition at 1000K and 20000 ML$s^{-1}$
  in top view (a) and 3D view (b). The gray scale shows the local
  height of the surface and the exposed part of the substrate is
  shaded in brown.
}
\end{figure}

We have simulated deposition of films with 8\% lattice misfit at
temperature 1000K and deposition rate 20000 ML$s^{-1}$.  Figure
\ref{F-D1000} shows the resulting morphology from a typical run at a
nominal film thickness of 3MLs. Isolated semi-spherical islands are
observed.  Most of them nucleate when the nominal coverage is about 1
ML and then grow steadily as more atoms are deposited. Coarsening via
exchange of atoms among islands (Ostwald ripening) also
occurs. Some small islands shrink and vanish
eventually.  However, coalescence of islands is suppressed by their
mutual elastic repulsion \cite {Jesson2004}. In fact, the edges of
neighboring islands are often deformed to avoid each others.

In our simulations, as in experiment, the deposition rate has a substantial effect on surface morphology. At the
rate considered above, island growth is limited by the supply of atoms.
Individual islands have already relaxed to their equilibrium shapes.
That is, deposition is slow relative to the formation dynamics and
geometrical relaxation of islands. In contrast, at lower deposition rates, we
observe that islands become larger and less dense because there is more
time for coarsening. For deposition faster than island
formation, layers of atoms quickly accumulate before the resulting
film roughens \cite{Lam}. With an abundant supply of atoms, we observe
that the roughening dynamics is  similar to that for annealing
except for a trivial vertical drift of the whole surface.
We will discuss annealing next.


\begin{figure}
\resizebox{0.42 \columnwidth}{!}{\includegraphics{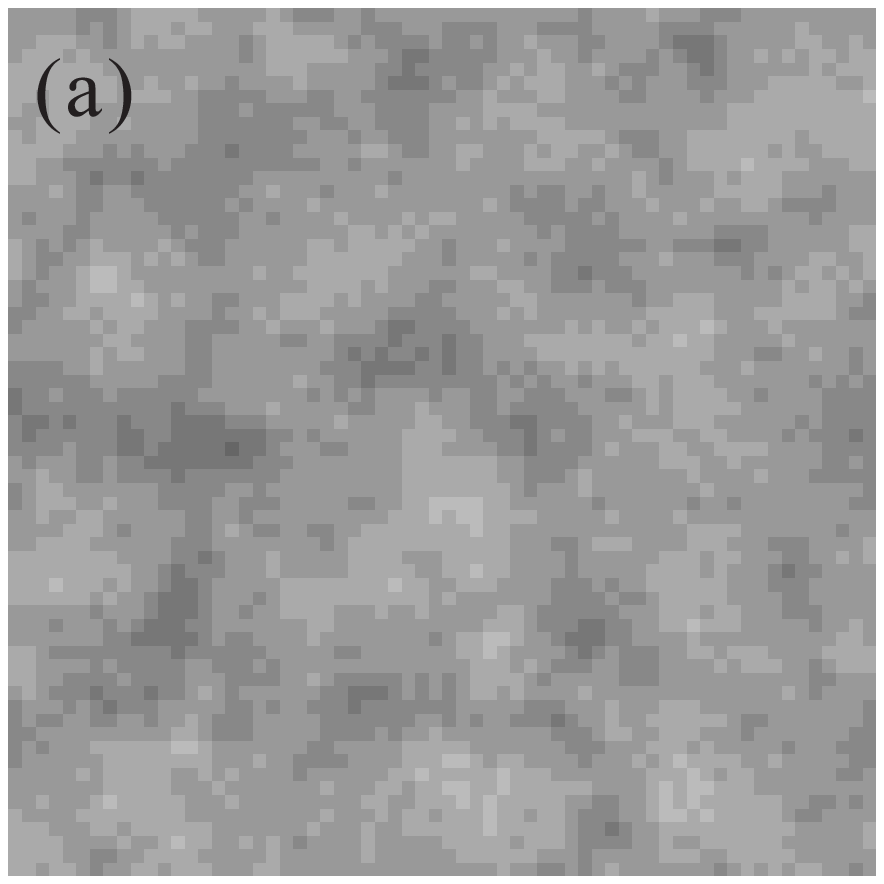}}
~
\resizebox{0.42 \columnwidth}{!}{\includegraphics{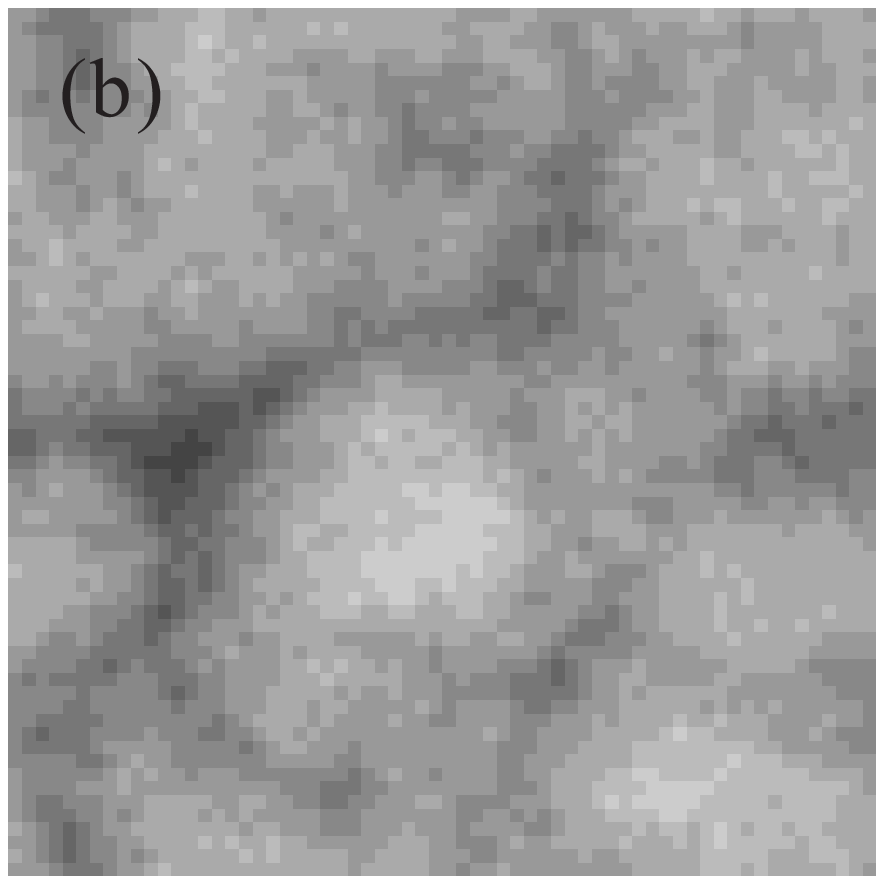}}
\\~\\
\resizebox{0.42 \columnwidth}{!}{\includegraphics{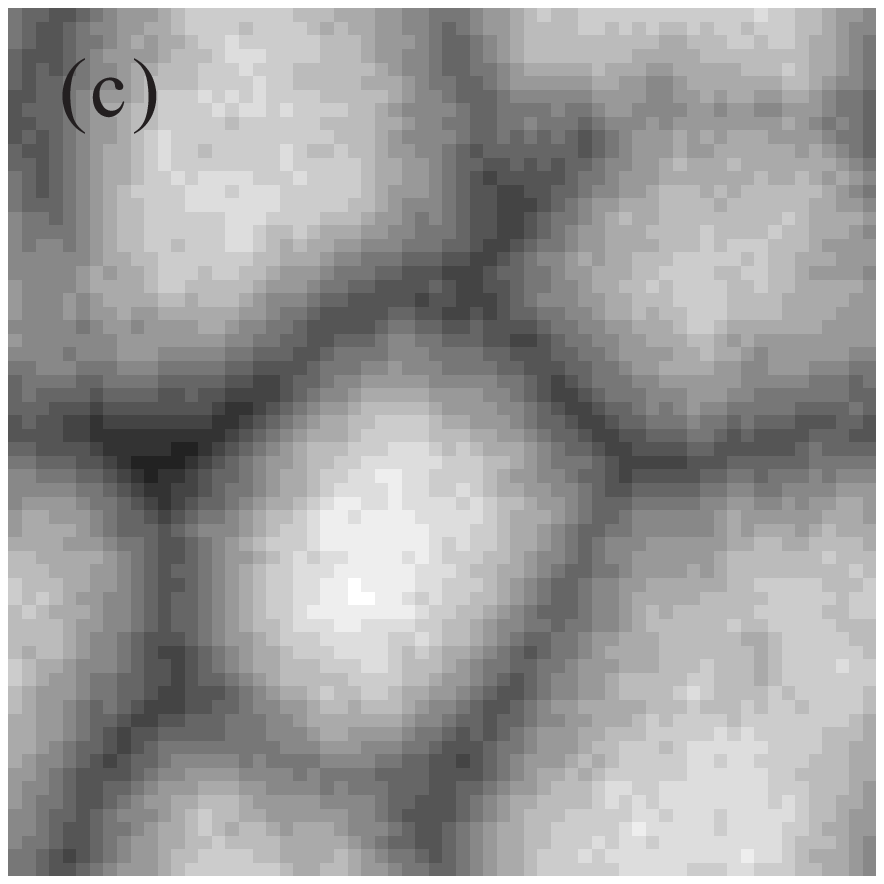}}
~
\resizebox{0.42 \columnwidth}{!}{\includegraphics{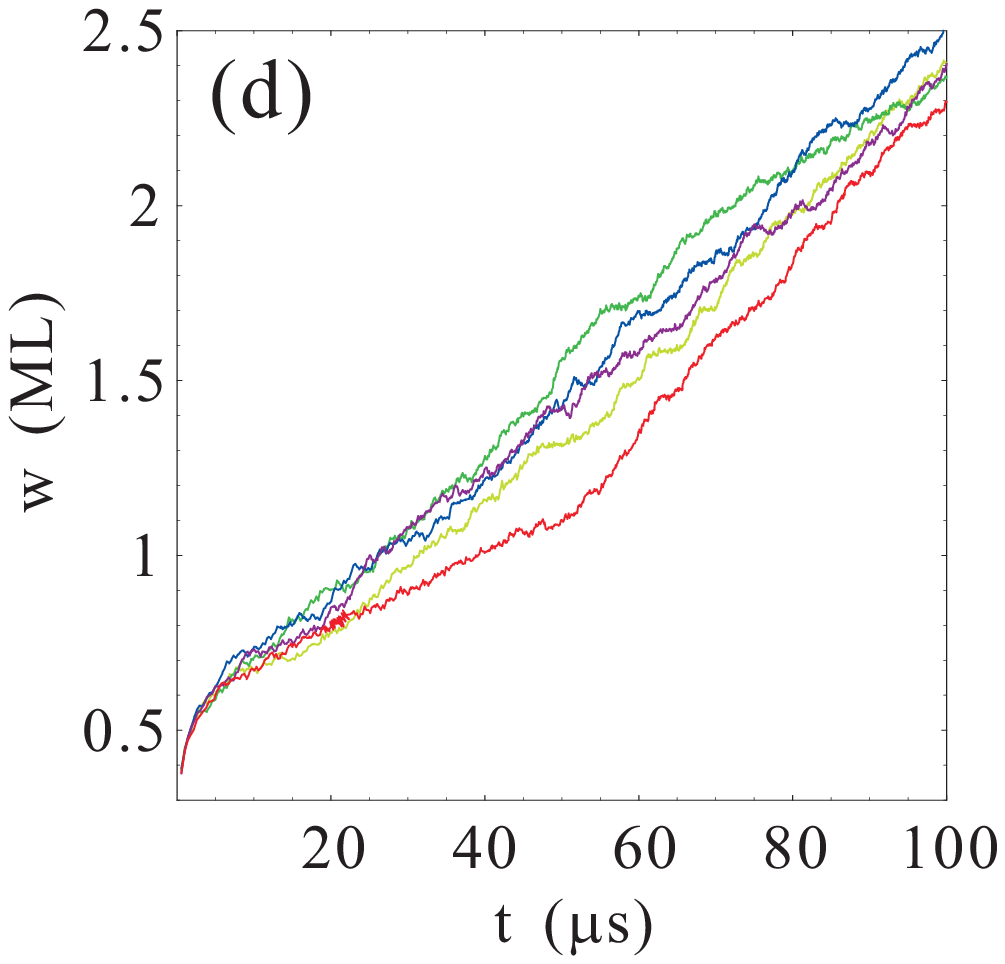}}
\caption{
  \label{F-A1000}
  Snapshots from annealing of an initially flat film at 1000K at
  time $t$ = 20 (a), 50 (b) and 100$\mu s$
  (c), and a plot of surface width $w$ against $t$ for 5
  independent runs (d).  }
\end{figure}

We have simulated
annealing of initially flat films with 10MLs of atoms and $6\%$
lattice misfit at 1000K. Figures
\ref{F-A1000}(a)-(c) show snapshots of the evolution.  2D islands and
pits first develop leading to a high step density [Fig.
\ref{F-A1000}(a)].  At this point, the film is still relatively flat and highly
stressed.  The misfit has little impact on the
morphology except for an enhancement of  the step density due to
a reduction of the
effective step free energy. As the roughness increases, long-range
elastic interactions begin to dominate and lead to the formation of 3D
islands and pits with gentle slopes [Fig.  \ref{F-A1000}(b)].
Subsequently, well developed 3D islands bounded by a network of
grooves emerge [Fig.  \ref{F-A1000}(c)]. Note that the surface
inclination at many grooves has reached its maximum value allowed in
our model. In experiments on the Si-Ge system, grooves are often bounded by 
[115] facets. The physical reasons for this might be the same, though our model
is too crude to select among facets.
 
The (100) surface studied above does not act like a true facet as is evident from
the abundance of surface steps in Fig. \ref{F-A1000}(a). This indicates that
1000K is above the surface roughening transition temperature. Thus, the
surface energy  varies smoothly with the
local inclination.  In this situation, the strain-induced roughening of an unfaceted
surface should be described by the
Asaro-Tiller-Grinfeld instability theory \cite{Grinfeld} which predicts
that random perturbations of the surface at sufficiently long
wavelength spontaneously amplify. The surface will gradually be
dominated by modulations at the most unstable wavelengths. 

Our annealing results at 1000K are consistent with the instability theory.
This is supported by a few characteristic features. First, the 
sidewalls of the newly emerging islands are gentle and their inclinations  
increase gradually rather than abruptly. Moreover, the island
base areas stay relatively constant. As a further evidence, Fig.
\ref{F-A1000}(d) plots the r.m.s. surface width $w$ against the
annealing time $t$ for 5 independent runs. We observe that $w$
increases steadily and the ensemble fluctuations are small as expected for
barrierless processes. The morphological development also
qualitatively resembles the initial evolution of
 Si$_{1-x}$Ge$_x$/Si(100) films at high temperature and low misfit
\cite{Floro}. Tersoff, et. al have argued that the  Si$_{1-x}$Ge$_x$(100) surface under
these conditions is not a true facet \cite{Tersoff2002} and
theories based on unfaceted surfaces should apply.


\begin{figure}
\resizebox{0.4 \columnwidth}{!}{\includegraphics{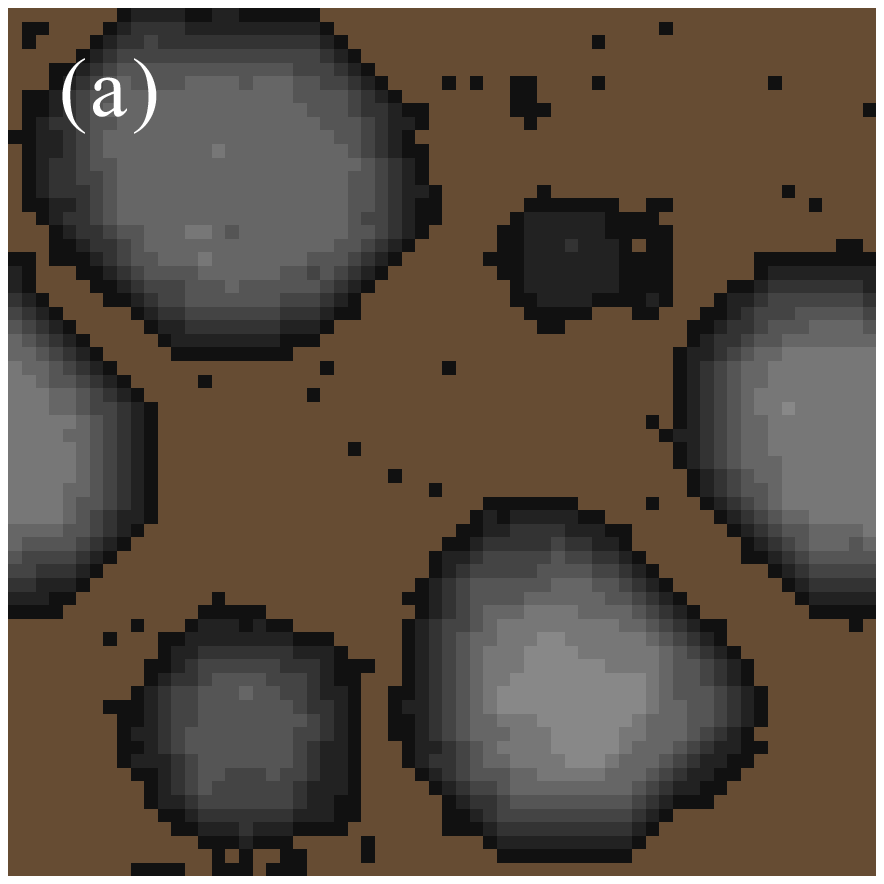}}
~
\resizebox{0.43 \columnwidth}{!}{\includegraphics{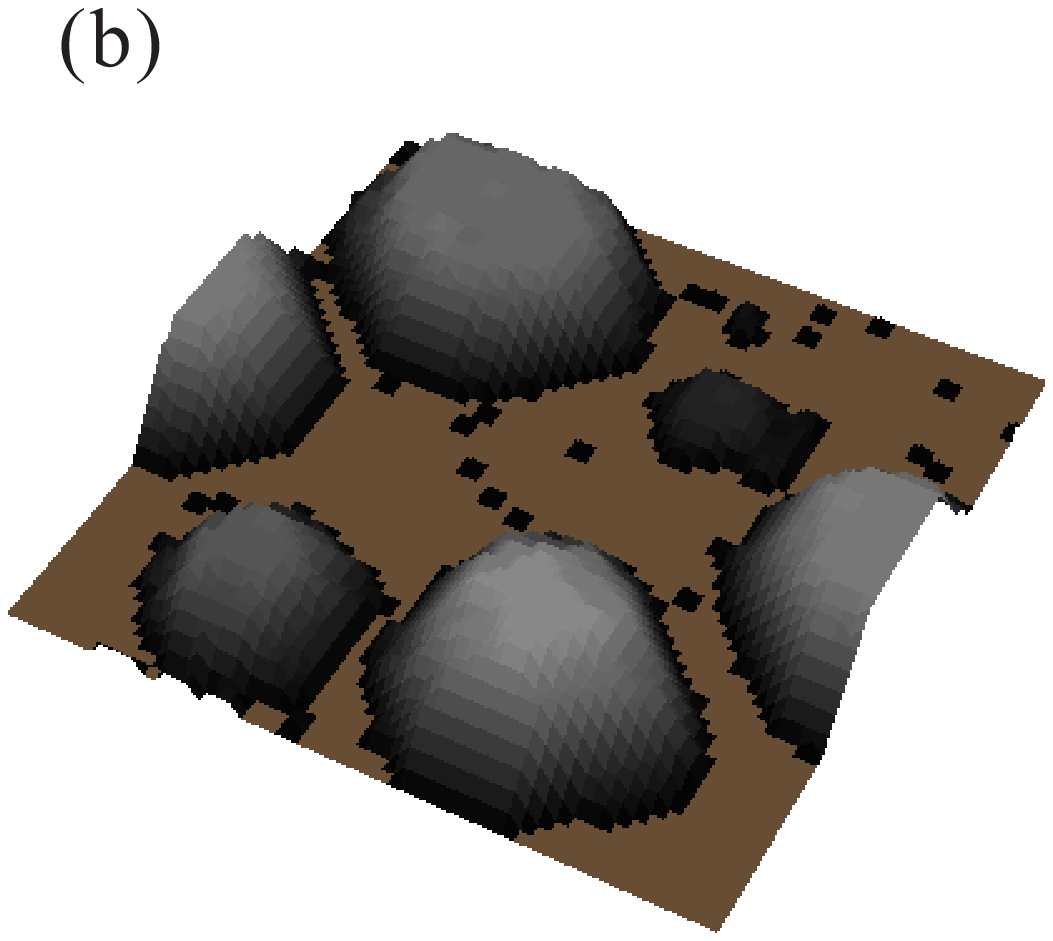}}
\caption{
  \label{F-D600}
  Surface from simulation of deposition at 600K and
  10 ML$s^{-1}$.  
}
\end{figure}

Next, we consider a lower temperature, 600K, which gives drastically
different morphologies indicating distinct roughening mechanisms.
Figure \ref{F-D600} shows a surface at a nominal coverage of 2MLs
from a simulation of deposition at 8\% misfit and 10 ML s$^{-1}$.  We
again observe isolated islands but they now take the shapes of
truncated cones. Most islands are out of equilibrium as their heights are
clearly limited by significant energy barriers for upper layer nucleation.


\begin{figure}
  \resizebox{0.42 \columnwidth}{!}{\includegraphics{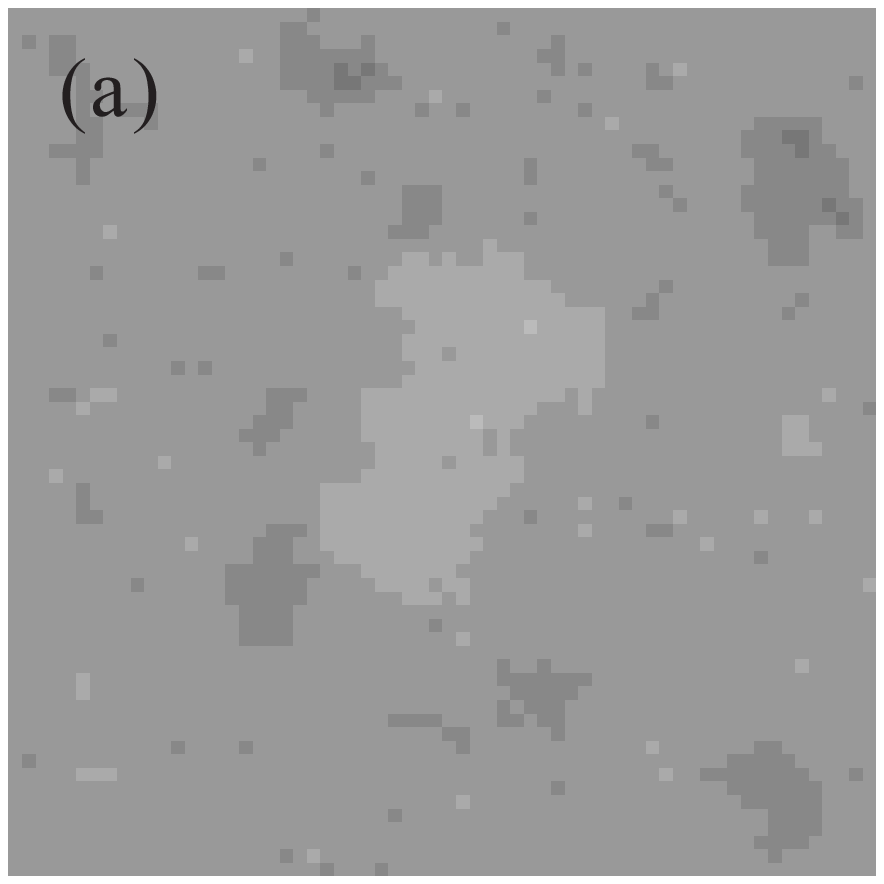}} ~
  \resizebox{0.42 \columnwidth}{!}{\includegraphics{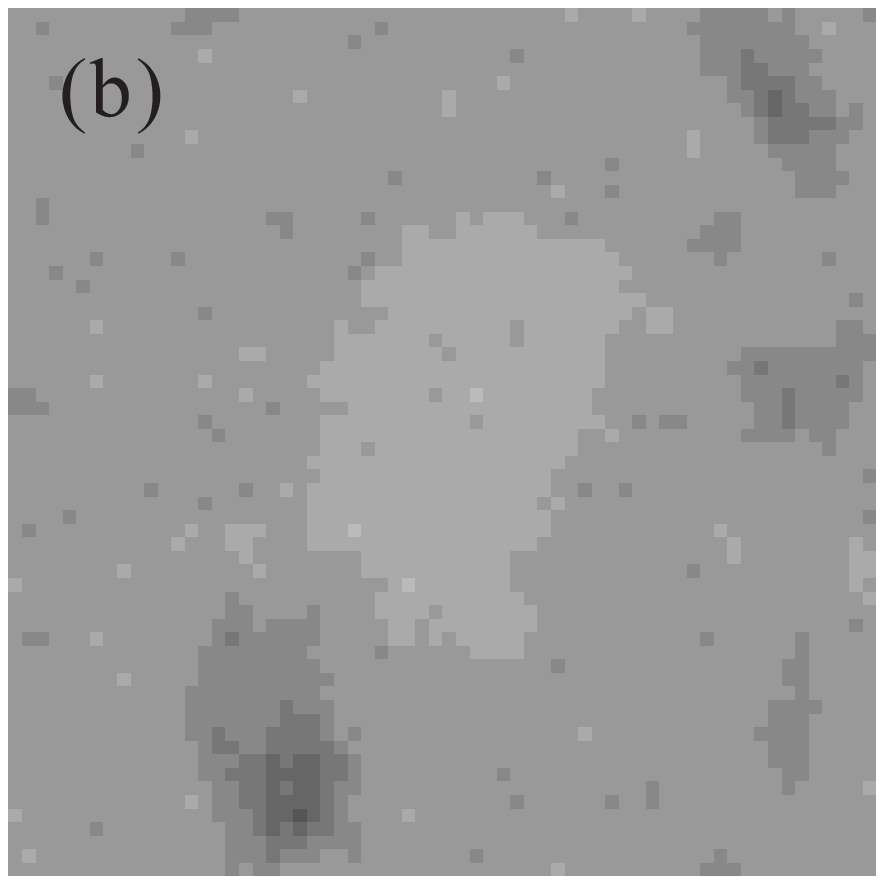}}
  \\~\\
  \resizebox{0.42 \columnwidth}{!}{\includegraphics{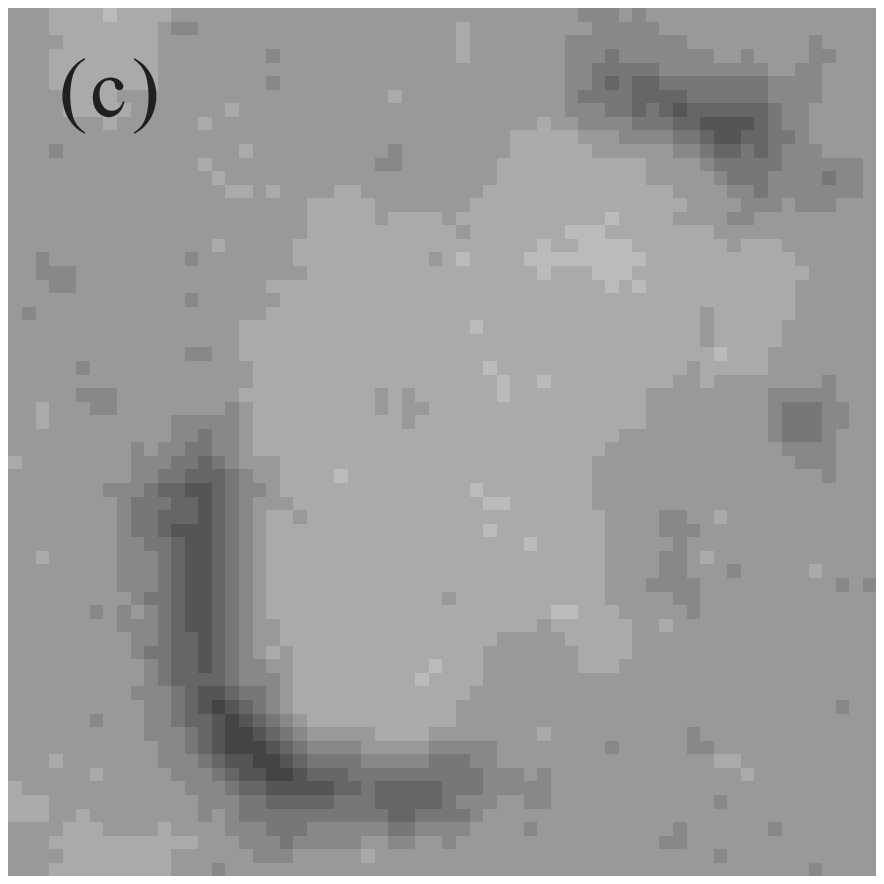}} ~
  \resizebox{0.42 \columnwidth}{!}{\includegraphics{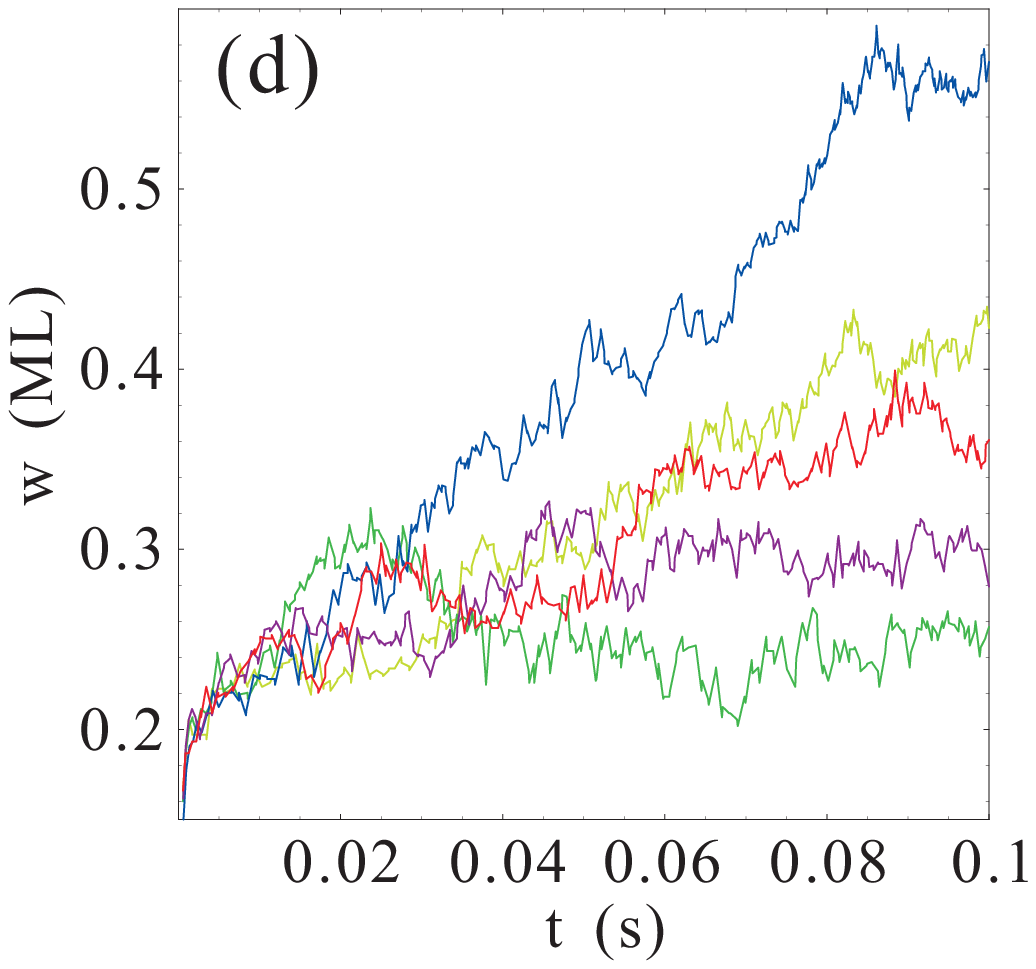}}
\caption{
  \label{F-A600}
  Snapshots from annealing of an initially flat film at 600K at
  time $t$ = 0.1 (a), 0.15 (b) and 0.22 $s$
  (c), and a plot of $w$ against $t$ for 5
  independent runs (d).  
}
\end{figure}

We have also simulated annealing at 600K. Figures
\ref{F-A600}(a)-(c) show three snapshots from a typical run. A large
2D island and a few smaller 2D pits first appear [Fig.
\ref{F-A600}(a)]. Later, 3D pits develop [Fig. \ref{F-A600}(b)].  They
then become increasingly eccentric and gradually turn into grooves
[Fig.  \ref{F-A600}(c)].  Analogous 3D structures
are also observed for deposition at rates fast
compared to roughening.

It is interesting to note that only part of the surface roughens even
after a long annealing time in sharp contrast to the high temperature
case. This strongly suggests that the surface is a true facet and 600K
is below the roughening temperature. The surface energy should be a
singular function of the slope and instability theory should not
apply. For this situation, a nucleation theory has been suggested for 3D
island or pit formation on faceted surfaces
\cite{Tersoff1994,Bouville}. According to this approach, an island or
pit has to overcome an energy barrier associated with a critical
volume before it can be stable.  Experiments on island formation at low
temperature and high misfit have indicated better agreement with
nucleation theory \cite{Tersoff1994}.

Figure \ref{F-A600}(d) plots the r.m.s. surface width $w$ against time
from 5 independent runs during early stage of roughening. There are
large ensemble fluctuations supporting the relevance of nucleation
processes. However, there exists no dominating jump in $w$ associated
with a single successful nucleation event after which $w$ grows
steadily.  Instead, multiple relatively rapid increments can be
observed and are associated with the creation of lower layers in the
dominant pits.  Conventional nucleation theory does not
successfully describe  the effects of the large barriers for nucleation
of further layers.  The formation of 3D pits in Figs.
\ref{F-A600}(a)-(c) and in fact also of the 3D islands in Fig.
\ref{F-D600} should best be described by a sequence of layer-by-layer
nucleation events. For a growing pit for instance, atoms are ejected
continuously while lateral expansion takes place at constant pit depth.
Once the bottom becomes sufficiently large, nucleation of a further
layer will be possible. The growth is thus based on the correlated
processes of continuous lateral expansion and periodic sudden
nucleation of deeper layers. The associated rates depend not only on
the pit geometry but also on the presence of nearby
islands or pits due to both exchange of atoms and elastic
interactions. Recent theories on the elasticity of step mounds should
be particularly relevant for further analysis \cite{Kaganer,Shenoy2002}.

The selection mechanism between islands and pits also deserves further
explanation.  Continuum elasticity theory shows that islands and pits
with infinitesimal slopes relieve elastic energy equally well
\cite{Tersoff1994}. It is visually apparent that an up-down
symmetry exists for surfaces in Fig. \ref{F-A1000}(a) and to a lesser
extent also in Fig.  \ref{F-A1000}(b). However, pits are increasingly
favored energetically compared to islands as local slopes become
steeper \cite{Vanderblit}. At low temperatures, the energy difference
is already significant for single layered structures. Specifically,
asymmetry between 2D islands and pits is already apparent in Fig.
\ref{F-A600}(a). There is typically one dominant island but a few
smaller pits. This is because the lower energy of pits also implies a
lower nucleation barrier. Pits can hence nucleate more quickly and are
more abundant. Because new islands are not nucleated, the existing one absorbs
all ejected atoms and grows quickly. Furthermore, the better stability
of pits also explains the development of 3D pits rather than 3D
islands in Figs.  \ref{F-A600}(b)-(c). We have observed 3D islands
only in Fig.  \ref{F-D600} for slow deposition. This is because under
this slow deposition rate, 3D islands are already able to develop before a
thick enough film can be formed to accommodate the pits \cite{Lam}.
Experimentally, the selected structure also turns from 3D islands to 3D
pits upon lowering the temperature and increasing the deposition rate
\cite{Gray2002}.

An interesting transition from pits to grooves is also observed in
Figs.  \ref{F-A600}(b)-(c). For shallow pits, a square base is
energetically preferred to a rectangular one \cite{Tersoff1994}. This
explains the more rounded shapes of the pits in Fig. \ref{F-A600}(b).
As the pits enlarge, their sidewalls also become steeper to relieve
the stress more efficiently. Grooves are then energetically preferred
to rounded pits because their linear extents are larger and can lead
to stress relief over a much wider region. Formation of grooves in
Fig.  \ref{F-A600}(c) is further enhanced by the stress around a 2D
island.  This is closely related to the phenomena of cooperative
nucleation \cite{Jesson1996} and also trench formation around 3D
islands \cite{Drucker}. The presence of a neighboring island however is
not essential as we have also observed pits turning into grooves far
away from any islands. Grooves are also observed in experiments from
annealing of pits \cite{Gray2004}.


In conclusion, we have applied a kinetic Monte Carlo method in 3D to
study morphological structures generated from deposition and annealing
of strained heteroepitaxy.  Under deposition conditions, morphologies
depend dramatically on whether deposition is slow compared to the
intrinsic roughening rate of the surface as in the 2D case \cite{Lam}.
For slow deposition,
isolated islands result and their formation and development are
limited by the supply of atoms. In contrast during fast deposition, 3D
structures form only after layers of atoms have accumulated and are
similar to those from annealing of initially flat
films. Morphologies from annealing further show strong dependence on
temperature which determines whether the initial surface is faceted.  
Upon annealing at high temperature, unfaceted surfaces develop
arrays of 3D islands via the Asaro-Tiller-Grinfield instability.  
In contrast, faceted surfaces
 at low temperature develop 3D pits via a
layer-by-layer nucleation mechanism. The pits later turn into grooves. 
We suggest that the
selection mechanisms between islands and pits as well as between pits
and grooves are of energetic origin.
Many of the general trends that we observe in our simulations are similar to experimental results. 


We thank J.A. Floro for helpful discussions. This work was supported
by HK RGC, Grant No. PolyU-5289/02P. LMS is supported in part by NSF grant No. DMS-0244419.

\end{document}